\documentclass[%
 reprint,
superscriptaddress,
 amsmath,amssymb,
 aps,
floatfix
]{revtex4-2}

\usepackage{graphicx}
\usepackage{dcolumn}
\usepackage{bm}

\usepackage[utf8]{inputenc}
\usepackage[none]{hyphenat}
\usepackage{tabularx}
\usepackage{amssymb,amsmath,bm}
\usepackage{graphicx}
\usepackage{enumerate}
\usepackage{cancel}
\usepackage{color}
\usepackage{float}
\usepackage[colorlinks=true,linkcolor=blue,urlcolor=blue,citecolor=blue]{hyperref}
\usepackage{caption}

\begin{document}

\preprint{APS/123-QED}

\title{A stochastic heat engine driven using a nonlinear protocol}

\author{Amrutayani Panda}
\email{ap23rs087@iiserkol.ac.in }
\affiliation{
 Department of Physical Sciences, Indian Institute of Science Research and Education, Kolkata, 741246 }%
 
\author{Biswajit Das}
 \email{biswajit.physics33@gmail.com}
 \thanks{Present address: Department of Physics, University of Konstanz, Germany}
\affiliation{
 Department of Physical Sciences, Indian Institute of Science Research and Education, Kolkata, 741246 }
 \author{Shuvojit Paul}
\affiliation{Department of Physics, Kandi Raj college, Kandi, Murshidabad, 742137}
 \author{Arnab Saha}
 \email{sahaarn@gmail.com}
  \affiliation{Department of Physics, University of Calcutta, 700009}
  \affiliation{Laboratoire de Physique Théorique et Modélisation, UMR 8089, CY Cergy Paris Université, 95302 Cergy-Pontoise, France}
 \author{Ayan Banerjee}
  \email{ayan@iiserkol.ac.in}
 \affiliation{
 Department of Physical Sciences, Indian Institute of Science Research and Education, Kolkata, 741246 }

\begin{abstract}

A colloidal particle confined in a time-dependent optical trap can function as a microscopic heat engine, with optimization strategies offering a route to enhanced performance. Here, we numerically investigate a Stirling heat engine operating in passive and active environments using a protocol inspired by Engineered Swift Equilibration (ESE). We analyze various fluctuating parameters throughout the cycle to validate the robustness of the engine, and demonstrate a significant enhancement in performance compared to conventional Stirling engines. Most crucially, we observe that the nonlinear protocol can even transform a heat pump-like operation into a genuine heat engine under strong activity, thereby surpassing bounds imposed on efficiency by high-temperature and quasi-static conditions. In contrast to the standard Stirling protocol, we observe that the nonlinear protocols appear to suppress the finite-time relaxation lag of the system from reaching an instantaneous equilibrium distribution, thereby improving engine performance at short cycle times. An analytical analysis of the variance dynamics where we quantify the relaxation lag between the actual and instantaneous equilibrium states clearly vindicates our understanding about the improved efficacy of the non-linear protocols. The proposed protocol is designed with experimental feasibility in mind, providing a practical route toward efficient microscopic heat engines.

\end{abstract}

\maketitle


\section{\label{sec:level1}Introduction}
The study of heat engines has undergone a profound transformation from the macroscopic, deterministic realm of classical thermodynamics to the highly fluctuating microscopic world of stochastic systems due to the advent of stochastic thermodynamics~\cite{bustamante2005nonequilibrium,jarzynski2012equalities,seifert2008stochastic, seifert2012stochastic}. Further, the advancement of micro-manipulation techniques has increased control over microscopic and nanoscopic systems and the precision of their measurements, helping the emergence of stochastic heat engines - devices that can harness random thermal motion at micro-scale - as both a fundamental research interest and technological frontiers~\cite{quinto2014microscopic,rossnagel2016single,blickle2012realization,martinez2016brownian}. As these systems become more advanced, there is a growing need to optimize them for enhanced performance to meet the demands of real-world applications. Importantly,
unlike macroscopic engines, stochastic heat engines~\cite{blickle2012realization,martinez2016brownian} work in an environment where fluctuations are not a mere perturbation but a dominant feature, and therefore have substantial effects in the corresponding thermodynamic parameters, such as the work or heat exchanged during each cycle~\cite{martinez2016brownian, manikandan2019efficiency, verley2014unlikely}. Such engines are often realized by holding colloidal particles in time-varying optical potentials ~\cite{blickle2012realization} due to their experimental accessibility and controllability.

Over the past decade, numerous theoretical and experimental~\cite{schmiedl2007efficiency,blickle2012realization,martinez2016brownian,krishnamurthy2016micrometre,rana2014single,holubec2020active,krishnamurthy2023overcoming,albay2023colloidal,chang2023stochastic,roy2021tuning,kwon2024effects,wiese2024modeling,holubec2014exactly,majumdar2022exactly} studies have explored Stirling engines using colloidal particles as the working substance in a passive thermal bath and active environments to study energy conversions at small scales. In active systems, activity has been introduced either through externally applied noise to mimic a bacterial bath or by using an actual bacterial suspension. Experimental implementations have commonly employed time-dependent optical traps operating under thermal and active noise conditions with tunable persistence. Now, it is well-known that the standard finite-time heat engine is bounded by the Carnot efficiency ($\eta_c$)~\cite{carnot1872reflexions} at the quasi-static limit and by the Curzon-Ahlborn efficiency~\cite{curzon1975efficiency} at maximum power. Moreover, for passive Stirling engines the quasistatic efficiency $\eta_q$ acts as the upper bound, while in active engines the limit is the high-temperature efficiency $\eta_\infty$~\cite{krishnamurthy2016micrometre}. Previous studies have shown that surpassing these limits is possible by introducing  a temperature dependence in the active noises~\cite{kwon2024effects} or by coupling the system to a stiffness-dependent noise ~\cite{albay2023colloidal}, so as to drive it further from equilibrium. Despite significant progress being made in understanding such systems, achieving high efficiency and power output simultaneously under finite-time conditions still remains a major challenge~\cite{krishnamurthy2023overcoming}. Importantly, conventional Stirling engines often follow linear protocols to complete the engine cycle which limits the flexibility of the process. Now, nonlinear processes often result in intriguing phenomena in physics, as is observed very commonly in optical processes \cite{boyd2008nonlinear} and electronics \cite{devices1976nonlinear} -- thus, an interesting question to ask is -- what would be the outcome with respect to efficiency and power of a microscopic heat engine if nonlinear protocols are utilized for the engine cycle? Motivated by this, we explore one such constructive example of a nonlinear protocol inspired by the Engineered Swift Equilibration (ESE) method~\cite{martinez2016engineered,chupeau2018thermal}, that is known to accelerate system relaxation, a factor which could be crucial for reducing irreversible work, so as to obtain higher output work from the system. Our aim is to test the hypothesis that -- along the same lines -- implementing a nonlinear protocol in an engine may reduce overall dissipation more efficiently, thereby enhancing performance even further under experimentally feasible conditions.

Accordingly, in this paper, we propose an approach for controlling the engine cycle by using a curved protocol as an alternative to the conventional linear one. Our protocol introduces a symmetric nonlinear trajectory that deviates from the traditional linear one, enhancing performance at short cycle times without compromising the stability of the engine which was confirmed from the fluctuations of stochastic quantities and the dissipation from the processes.
Although this protocol is not derived from a formal optimization process, it demonstrates significant improvement in efficiency and power relative to conventional methods. Importantly, it remains experimentally realizable, as it accounts for the interdependence of effective temperature and trap stiffness, something that is often overlooked in theoretical optimization processes~\cite{chatterjee2025optimal,schmiedl2007optimal,schmiedl2007efficiency,PhysRevE.98.022133,ye2022optimal} that assume temperature to be an independent parameter. However, this assumption ignores the fact that for a colloidal particle in harmonic confinement, the instantaneous effective temperature $T_{eff}(t) \sim k(t)\langle x^2(t)\rangle /k_B$, where $k_B$ is the Boltzmann constant, will vary with the strength of confinement ($k$). Thus, in our study, we perturb the working substance of the micro heat engine, i.e. the colloidal particle, is perturbed externally using white noise for passive systems (to mimic the state of higher effective temperatures), and Ornstein-Uhlenbeck noise for active systems~\cite{martinez2013effective}, effectively tuning the non-equilibrium fluctuations to achieve both activity and higher effective temperature. We believe that our work would have a significant contribution to the growing research in this area by demonstrating a practical, experimentally viable protocol that enhances performance through nonlinear driving, even in the presence of strong fluctuations and non-equilibrium activity.

\section{Protocol}

\begin{figure}
    \centering
    \includegraphics[width=1\linewidth]{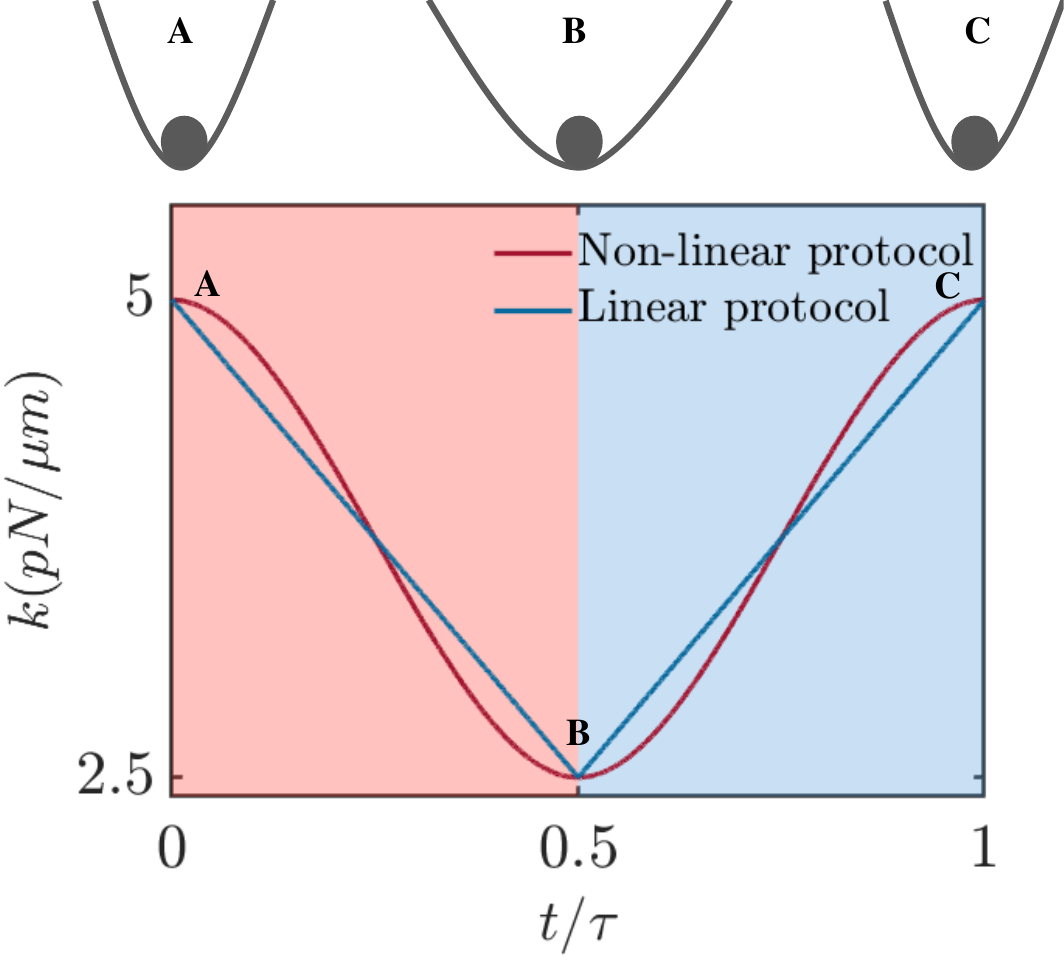}
    \caption{\raggedright Time-dependent Stiffness profile for both protocols}
    \label{fig:1}
\end{figure}

We consider the engine to operate in a two-step process
1. Isothermal expansion at higher effective temperature, and
2. Isothermal compression at lower (room) temperature (Fig.~\ref{fig:1}).
The temperature jumps instantaneously between compression and expansion. As usual, the expansion and compression steps of the engine can be mimicked through the time-varying stiffness of the optical tweezers. The features of a microscopic engine, as carried out in several experimental and numerical realizations \cite{blickle2012realization,krishnamurthy2016micrometre,albay2023colloidal,roy2021tuning,zakine2017stochastic,krishnamurthy2023overcoming,chang2023stochastic,holubec2020active,kwon2024effects,rana2014single,wiese2024modeling}, can be achieved straightforwardly by following a linear variation of the stiffness parameter ($k(t)$). In this context, the variation of $k(t)$ over the cycle time $\tau$ can be considered as, 
\begin{equation}
\label{eq_linear_kt}
    k(t)=\begin{cases}
        k_{max}-\frac{2}{\tau}(k_{max}-k_{min})t,&  0\leq t\leq \frac{\tau}{2}\\
        k_{min}-(k_{max}-k_{min})(1-\frac{2t}{\tau}),&  \frac{\tau}{2}< t\leq \tau
    \end{cases}
\end{equation}
which we term as \textit{linear protocol} with $k_{min}$ and $k_{max}$ being the minimum and maximum stiffness constants, respectively. 

Since we are particularly interested in the performance of an engine following a nonlinear variation of the stiffness constant over the cycle, we design a protocol where the $k(t)$ varies nonlinearly throughout the engine cycle. This particular variation of $k(t)$, termed as \textit{nonlinear protocol}, can be explicitly written as, 
\begin{equation}
\label{eq_nonlinear_kt}
    k(s)=\begin{cases}
        k_{max}-\Delta k (3s^2-2s^3),&  0\leq s\leq 1\\
        k_{max}-\Delta k (3(2-s)^2-2(2-s)^3
        ),&  1< s\leq 2
    \end{cases}
\end{equation}
where, $\tau$ is the cycle time, $\Delta k=(k_{max}-k_{min})$ and $s=\frac{2t}{\tau}$. We fix $k_{max}=5pN/\mu m$ and $k_{min}=2.5pN/\mu m$ throughout our study.
Note that, this non-linear protocol is inspired by the central idea of the ESE protocol~\cite{martinez2016engineered}: employing nonlinear, time-dependent control of the trap parameters to suppress lag and irreversibility during finite-time transformations. Unlike the original ESE formulation, which is designed to exactly connect two equilibrium states in a fixed time, our protocol is tailored to cyclic engine operation.

In both protocols, the compression phase is kept isothermal at room temperature, which is the same as the bath temperature. However, to mimic the effect of a higher temperature during expansion, a stochastic force is applied to the minima of the potential, thereby effectively increasing its kinetic temperature without physically raising the temperature of the surrounding bath. Further details about the application of the external drive are discussed later on. 

\subsection*{Energetics}
To assess the performance of a microscopic heat engine, trajectory energetics, especially the work done and the heat over a cycle, can be estimated using Sekimoto's technique~\cite{sekimoto1998langevin}. 
Since the engine is a thermodynamic cycle consisting of a single colloidal particle subjected to a time-dependent optical trap experiencing a harmonic $\frac{1}{2}k(t)x(t)^2$
potential, with time-dependent stiffness $k(t)$, 
the work done ($W$) and heat ($Q$) over the full cycle (in the overdamped regime) given by

\begin{equation}
    W=\frac{1}{2}\int_{0}^{\tau} dt\ x^2(t)\circ\ \dot k(t), \\
\end{equation}
\begin{equation}
    Q=\int_{0}^{\tau}\ dt \Bigl(k(t)\cdot x(t)\Bigl)\circ\ \dot x(t).
\end{equation}
Here, $\circ$ denotes a \textit{Stratanovich} product. Furthermore, we can estimate power as the average work extracted per unit time: $P=-\frac{\langle W\rangle}{\tau} $, with $\tau$ being the total cycle time. The efficiency of the engine can be computed as $\eta=-\frac{\langle W\rangle}{\langle Q_h\rangle}$, which is the total average work extracted upon average heat absorbed. Here, $Q_{h}$ is total heat absorbed by the system in the first half ($0$ to $\frac{\tau}{2}$) of the cycle. By convention, negative work is considered as the work done by the system, whereas positive heat corresponds to the heat absorbed from the environment.

Now, to evaluate the efficiency and power of our engine setup, we numerically integrate the corresponding overdamped Langevin equation with time step $dt=0.0001s$, which is sufficiently less than all other timescales of the system. The results are then averaged over an ensemble of 500 trajectories, each performing 1000 cycles to ensure statistical significance. 

With the framework of protocols and energetics established, we proceed to numerically test the engine in both passive and active environments. This enables us to evaluate and compare the effect of the driving protocol's shape on the performance of an engine.
\begin{figure}
    \centering
    \includegraphics[width=1\linewidth]{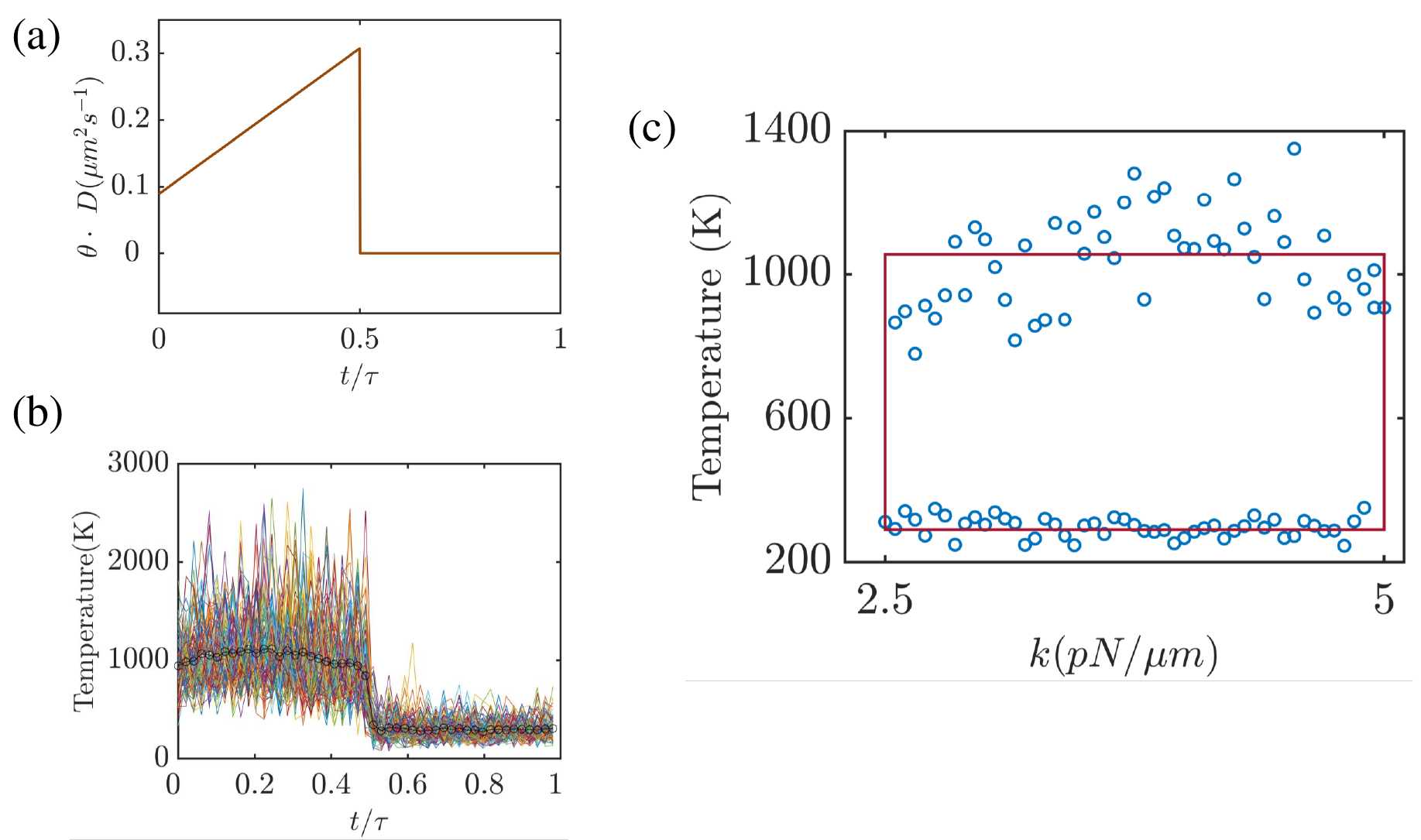}
    \caption{\raggedright (a) Variation of strength of the external noise throughout the engine cycle. (b)Temperature profile corresponding to the passive engine, (c)\ $k(t) $ vs $ T_{eff}(t)$ plot of the Passive Stirling cycle under quasi-static conditions.}
    \label{fig:2}
\end{figure}
\section{Passive Engine}

Let us now consider the colloidal particle to be trapped in a harmonic potential and immersed in a passive environment, such as water. In the overdamped regime, the dynamics of the particle is governed by the overdamped Langevin equation. The effective temperature of the engine is modulated by externally perturbing the mean position of the colloidal particle using an added white noise. This external noise is applied only during the first half of the engine cycle, effectively raising the effective temperature of the system. In the second half of the cycle, the temperature is maintained at room temperature, which corresponds to a passive thermal environment. The dynamics of the system is governed by the following equation:
\begin{equation}
     \dot x(t) = -\frac{k(t)}{\gamma}\Bigl(x(t)-\lambda (t)\Bigr)+\sqrt{2D}\ \xi(t)
\end{equation} 
where, $\gamma$ is the friction coefficient of the surrounding medium, and $\xi (t)$ is the white noise with correlation $\langle \xi(t) \xi(t') \rangle = \delta(t-t')$. Here,  $D=\frac{k_B T}{\gamma}$ is the diffusion coefficient with T being the room temperature 295 K. 
The term $\lambda(t)$ denotes the position of the trap center under externally applied noise. In the numerical implementation, for the passive engine scenario, the trap center position $\lambda_i$ at time step ($i$) is sampled as a discrete random variable: $\lambda_i = \sqrt{2\theta D}\ \eta_i, \quad \eta_i \sim \mathcal{N}(0,1)$, where $\eta_i$ is a standard normal random variable. This choice imposes a discrete trap-displacement variance of $\langle \lambda_i^2 \rangle = 2\theta D$ at each integration step $dt$. The parameter $\theta$ serves as a scaling factor that controls the amplitude of the external trap fluctuations, consequently, modulates the effective temperature of the system.

To achieve a higher effective temperature of approximately 1000 K, the amplitude of the applied noise is carefully tuned. Since the effective temperature depends on the stiffness of the trap, the noise amplitude is varied in a direction opposite to that of the stiffness as shown in Fig.~\ref{fig:2}(a). This noise and stiffness counterbalance each other, resulting in a nearly constant effective temperature throughout the process (Fig.~\ref{fig:2}(b)). The value of $\theta$ is manually tuned for quasi-static time scales through multiple iterations of trial and error, and is varied in the range of approximately 0.55 to 1.9 for the linear protocol and 0.5 to 1.9 for the non-linear protocol. This adjustment allowed fine control over the effective temperature of the system during the process of expansion of the engine cycle, enabling a target temperature of around 1000 K to be achieved (Fig.~\ref{fig:2}).

\begin{equation}
    T_{eff}(t)=\begin{cases}
        T_h\approx 1000\ K,&  0\leq t< \frac{\tau}{2}\\
        T=T_c=295\ K,&  \frac{\tau}{2}< t \leq \tau
    \end{cases}
\end{equation}

Note that,  the fluctuations in $T_{eff}$ arise due to the combined effect of thermal noise and the externally applied noise $\lambda (t)$ which is itself also stochastic. In our system, the controlled noise enters multiplicatively through the time-dependent trap stiffness $k(t)$ (the time-dependent stiffness $k(t)$ scales the strength of the applied noise), which leads to additional modulation of fluctuations in the instantaneous effective temperature. However we would like to emphasize that we are mainly concerned with the mean $T_{eff}$ over each stage of the cycle. These mean $T_{eff}$ values show significantly less fluctuations (over multiple runs with a fixed cycle time of the engine) compared to the instantaneous ones, and they remain near the desired 1000K temperature throughout the cycle. This ensures that the engine operates under well defined thermodynamic conditions when averaged over ensemble realizations.

All parameter values used throughout the analysis are experimentally feasible and consistent with existing experimental studies ~\cite{blickle2012realization,albay2023colloidal,roy2021tuning}. Moreover, the method we employ to maintain an effectively elevated temperature while varying the trap stiffness is grounded in established experimental practice. In practice, this can be achieved by passing the laser beam (used for trapping the microparticle) through an acousto-optic deflector (AOD) and tuning both the amplitude and the frequency of the radio-frequency (RF) drive~\cite{manikandan2021quantitative, das2023enhanced}. This simultaneous modulation adjusts the intensity and varies the position of the output beam (external fluctuating force) in a controlled manner, thereby mimicking the desired effective temperature at different trap stiffness values. Alternatively, electrical noise can be introduced through external electrodes attached to the sample chamber to generate a controllable fluctuating force on the trapped particle~\cite{saha2023information, martinez2013effective}, while the trapping-beam intensity is simultaneously  tuned by the AOD in a precise manner.

\begin{figure}
    \centering
    \includegraphics[width=1\linewidth]{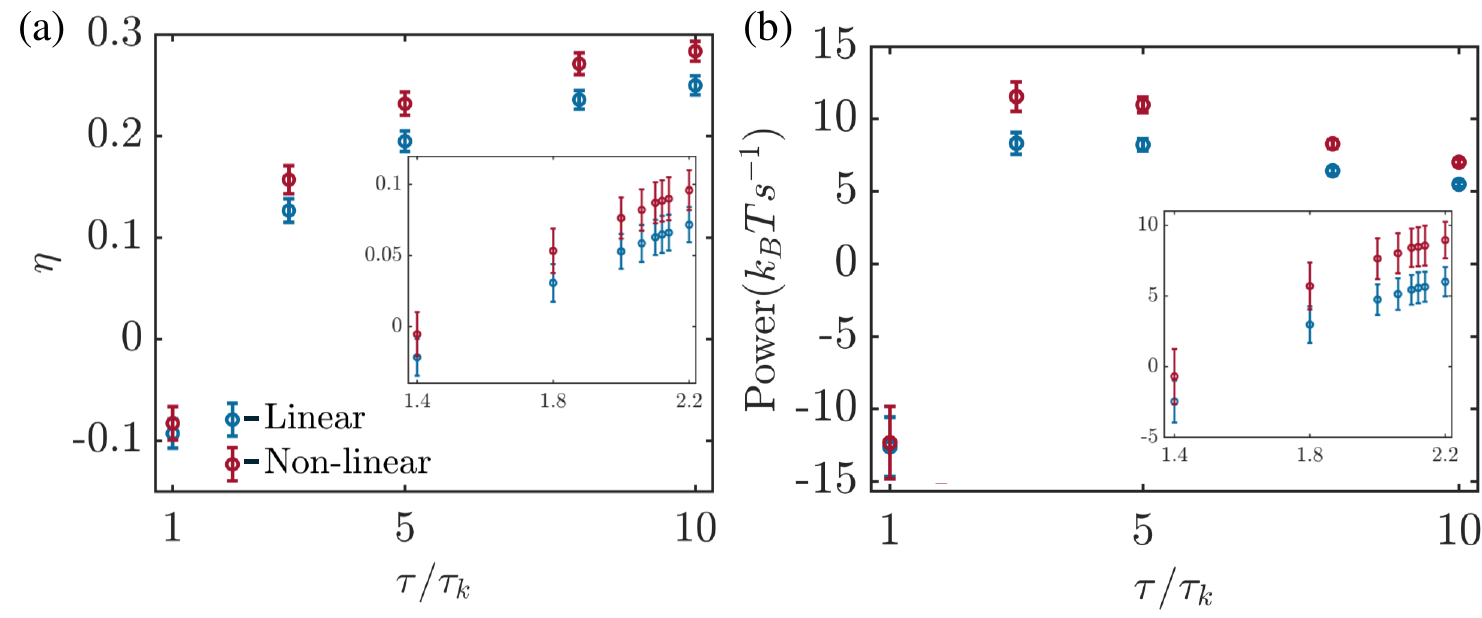}
    \caption{\raggedright Performance plots of the passive engine. (a) Efficiency as a function of short cycle times (inset: Efficiency for cycle times very close to $\tau_k$), (b) Power output at short cycle times (inset: Power trend for cycle durations near $\tau_k$)}
    \label{fig:3}
\end{figure}

 \begin{figure}
     \centering
     \includegraphics[width=1\linewidth]{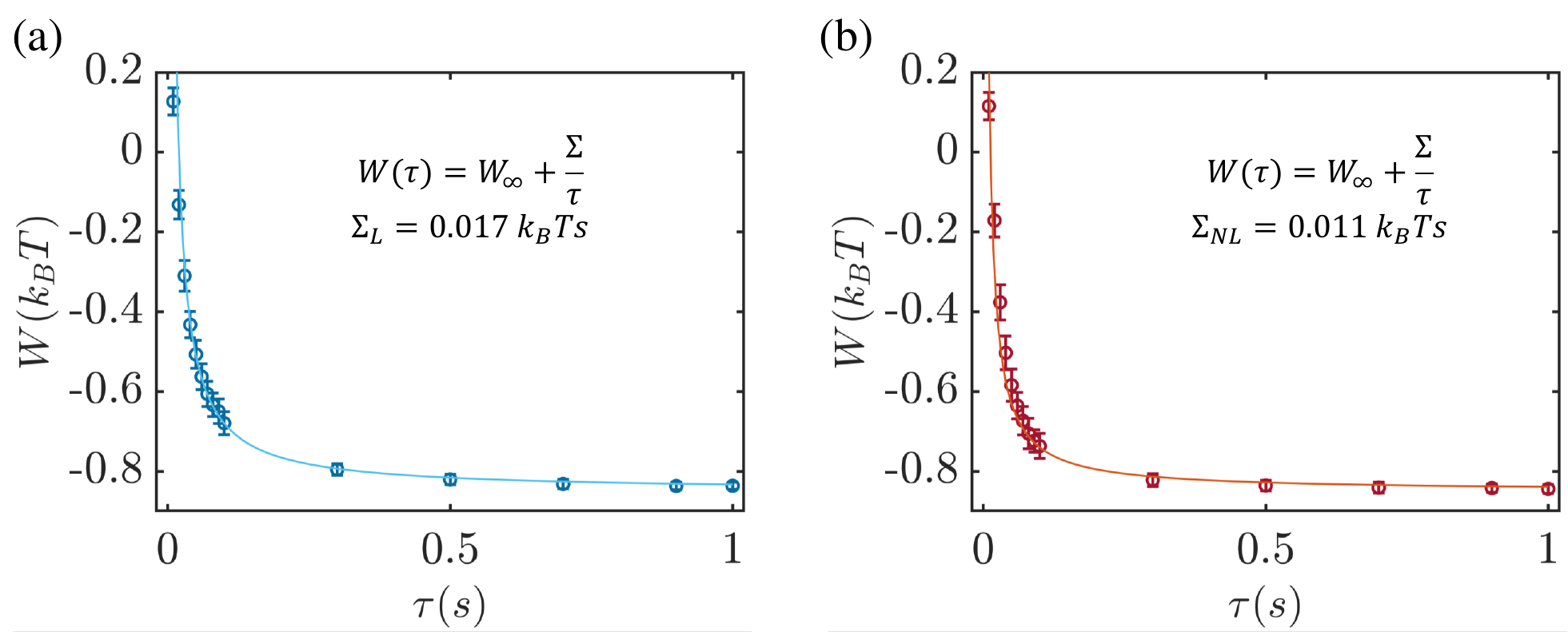}
     \caption{\raggedright Mean work done in the passive engine under both (a) linear and (b) nonlinear protocols. A strong agreement is observed between the analytical expression and simulation data, with fit parameters yielding $0.017\ k_B Ts$ for the linear protocol and $0.011\ k_B Ts$ for the nonlinear protocol.}
     \label{fig:4}
 \end{figure}

\subsection*{Performance of passive engine}

\begin{figure*}
    \centering
    \includegraphics[width=1\linewidth]{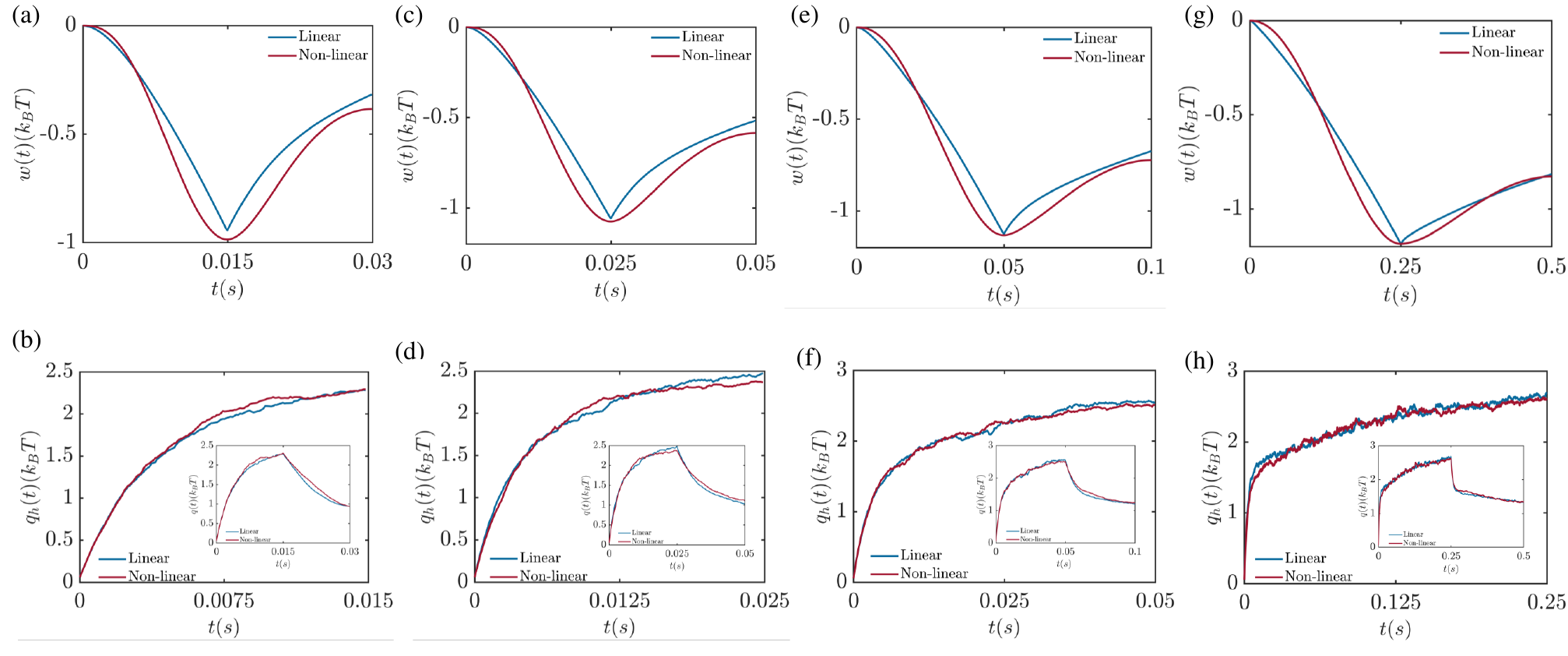}
    \caption{\raggedright Cumulative work accumulated over the whole duration of the engine cycle for different cycle times, (a)$\tau=0.03s=3 \tau_k$ (c)$\tau=0.05s=5 \tau_k$ (e)$\tau=0.1s=10 \tau_k$ (g)$\tau=0.5s=50 \tau_k$. Corresponding plots of cumulative absorbed heat during the engine operation (inset: Cumulative total heat exchanged throughout the engine cycle), (b)$\tau=0.03s=3 \tau_k$ (d)$\tau=0.05s=5 \tau_k$ (f)$\tau=0.1s=10 \tau_k$ (h)$\tau=0.5s=50 \tau_k$.}
    \label{fig:5}
\end{figure*}
To compare the performance of the engines operating under both protocols, we first compute efficiency and power as a function of cycle time. We observed a significant increase in both parameters, efficiency and power, at short time regimes for the nonlinear protocol (Fig.~\ref{fig:3}). The cycle times for the performance plots are expressed in terms of the relaxation time scale $\tau_k = {\gamma}/k_{min}=0.01s$ for better visualization. This enhancement in efficiency and power is consistent over all cycle times. 

To understand the underlying mechanism of this enhancement, we next focus on the work dissipated throughout the cycle following a particular protocol. In this regard, the mean work done over a finite cycle time can be expressed analytically ~\cite{esposito2010efficiency,schmiedl2007efficiency,sekimoto1997complementarity,bonancca2014optimal} as $W(\tau)=W_{\infty}+W_{diss} \equiv W_{\infty}+\Sigma/\tau$ where $\Sigma$ is the irreversibility parameter, also known as Sekimoto-Sasa constant. We then fitted this expression to the mean work done of the engines following our protocols as shown in Fig.~\ref{fig:4}, and found that the irreversibility parameter is lower for the nonlinear protocol ($\Sigma_{NL}=0.011\ k_B Ts$) compared to the linear protocol ($\Sigma_{L}=0.017\ k_B Ts$). The enhanced performance of the nonlinear protocol has a clear physical origin. In finite-time Stirling cycles, dissipation is largely caused by the system’s inability to instantaneously follow rapid changes in trap stiffness, resulting in a delayed response towards the instantaneous equilibrium distribution. By shaping the driving protocol nonlinearly, this delay is reduced, leading to lower dissipated work and improved engine performance. This interpretation is further supported by an analytical treatment of the variance dynamics, which quantifies the relaxation lag between the actual and instantaneous equilibrium states of the system. The analysis demonstrates that the nonlinear protocol systematically suppresses this lag, thereby reducing finite-time irreversibility. A detailed derivation together with the analytical lag calculations and the average work extracted by the engine is provided in Appendix C. Overall, the nonlinear protocol reduces the dissipated work by approximately $35\%$ compared to the linear, apparently because the corresponding drive matches the trap's relaxation dynamics, cutting down sharp bursts of dissipation~\cite{schmiedl2007efficiency}. This improvement is also evident from the cumulative plots representing the total energy accumulated up to time $t$, obtained by integrating the instantaneous work and heat along each stochastic trajectory (Fig.~\ref{fig:5}). The results averaged over 200 trajectories each performing 500 cycles show that the nonlinear protocol yields a higher amount of extracted work compared to the standard linear protocol.  Although the cumulative absorbed heat remains nearly the same for both protocols, this difference in work output contributes to the overall increase in efficiency. Note that, near the quasi-static limit (at long cycle times), the efficiency corresponding to both the protocols tends to saturate to an efficiency $\eta_{q} = \frac{\eta_c}{1+\frac{\eta_c}{\ln(\frac{k_{max}}{k_{min}})}}=0.33$~\cite{blickle2012realization}.
Here, $\eta_c$ denotes the Carnot efficiency.

Nevertheless, the passive engine operating under the nonlinear protocol is found to outperform the one that follows the usual linear protocol. This naturally leads us to the next plausible scenario: an active engine, where the external drive itself may possess an intrinsic relaxation timescale. 

\section{Active Engine}

For an engine operating in an active bath namely, systems characterized by finite persistence times, such as bacterial suspensions or by non-equilibrium fluctuations generated by self propelled particles~\cite{wu2000particle,han2017effective,sanchez2012spontaneous,ramaswamy2017active,bechinger2016active,ortlieb2019statistics,PhysRevLett.117.038103}, the dynamics can be modeled by the following set of coupled equations:
\begin{equation}
\label{eq:active_engine}
    \dot x(t)=-\frac{k(t)}{\gamma}\Bigl(x(t)-\lambda(t)\Bigr)+\sqrt{2D}\ \xi(t)
\end{equation}
and,
\begin{equation}
\label{eq:active_drive}
    \dot\lambda(t)=-\frac{\lambda(t)}{\tau_o}+\frac{\sqrt{2D_e}}{\tau_o}\ \zeta(t),
\end{equation}
with $\langle \xi(t) \rangle = \langle \zeta(t)\rangle = 0$, $\langle \xi(t)\xi(t') \rangle = \langle \zeta(t)\zeta(t') \rangle = \delta(t-t')$ and $\langle \xi(t)\zeta(t') \rangle = 0$.  
Here $\lambda(t)$ is the Ornstein-Uhlenbeck (OU) noise that can be applied externally to mimic the activity of the bath.  This noise is exponentially correlated  $\langle\lambda(t)\lambda(t')\rangle=\frac{D_e}{\tau_o}\exp(-\frac{|t-t'|}{\tau_o})$, with $\tau_o$ being its correlation (or persistence) timescale~\cite{maggi2014generalized,chaki2019effects,das2023enhanced,saha2019stochastic}.
The noise strength is defined as $\frac{D_e}{\tau_o^2}=A(k)\cdot D$ where $D=\frac{k_BT}{\gamma}$ and $A(k)$ is the stiffness-dependent amplitude. As mentioned earlier, this active noise mimics the behaviour of a bacterial bath and increases the effective temperature of the system. To design the cyclic engine, the noise is applied only in the first half of the cycle, similar to the passive scenario. 

\begin{figure}
    \centering
    \includegraphics[width=1\linewidth]{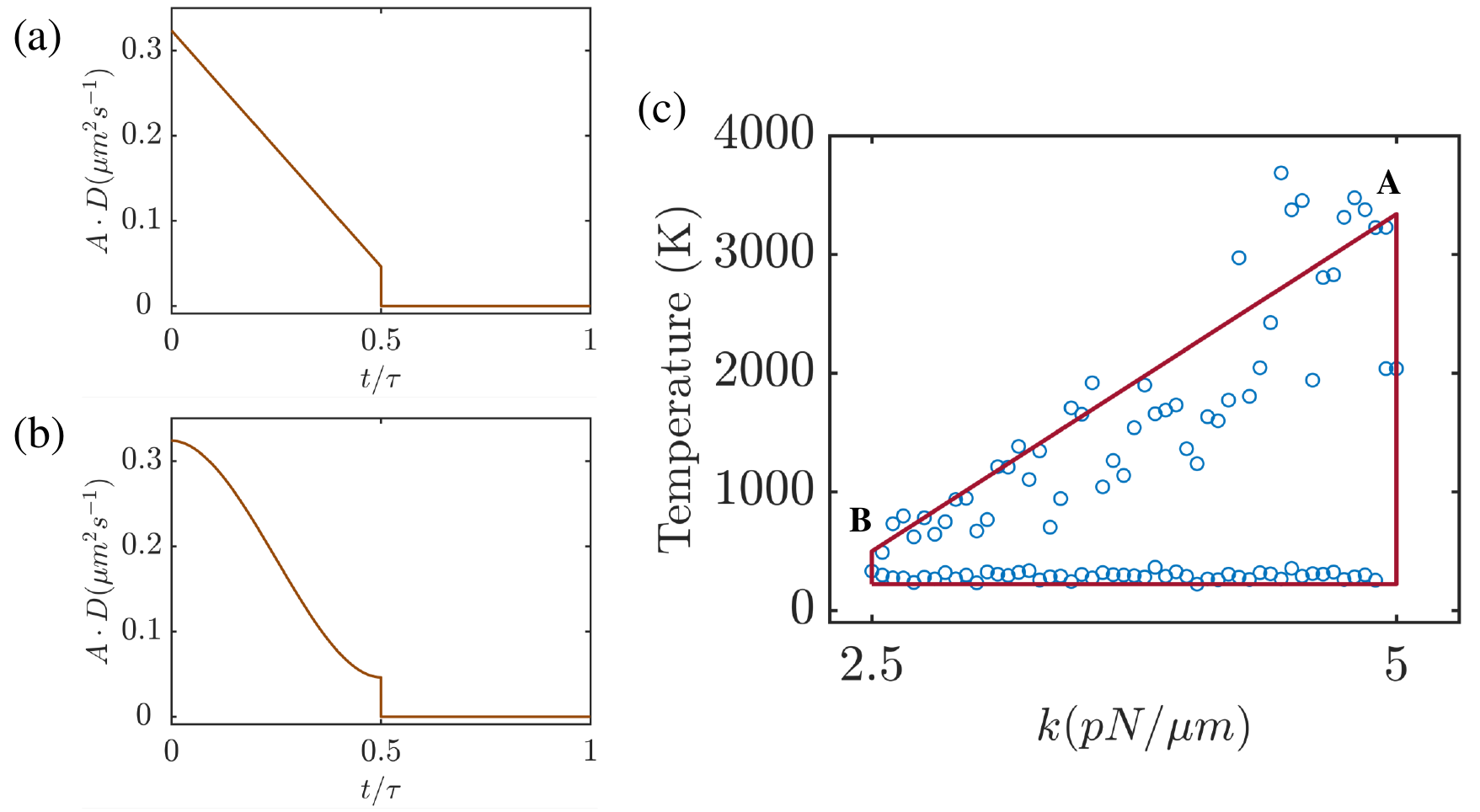}
    \caption{\raggedright Stiffness-dependent noise strength profile for both (a) linear and (b) nonlinear protocols. (c) $ k(t)$  vs  $T_{eff}$ diagram for an active engine.}
    \label{fig:6}
\end{figure}

Since the effect of active noise on the trapped particle will depend on the strength of confinement, the effective noise strength would be stiffness-dependent. To model this, we consider the amplitude of the correlated noise to follow the expression: $A(k)=T_A+a(k-k_{max})$ where $T_A$ and $a$ are the control parameters. The combination of this noise strength and room temperature ($T$) then gives the effective temperature~\cite{zakine2017stochastic,maggi2014generalized,martinez2013effective}.
\begin{equation}
    T_{eff}=T+\frac{A(k)\cdot D\cdot \tau_{o}^2}{k_{B}(\gamma+k\cdot \tau_{o})}.
\end{equation}
For our engine, the control parameters $T_A$ and $\frac{a}{a_c}$ are tuned such that the effective temperature begins approximately around $1000K$, with $\frac{a}{a_{c}}=2$, $T_A=2$. The critical value $a_{c}=\frac{T_{A}\cdot \tau_{o}}{\gamma+k_{max}\cdot \tau_{o}}$, corresponds to the condition $T_{eff}^A=T_{eff}^B$, where the effective temperatures at points A and B become equal, effectively making it an isothermal step similar to that of a passive engine. This methodology follows the approach described in Ref.~\cite{albay2023colloidal}. Fig.~\ref{fig:6}(a),(b) illustrate the noise profile for both protocols, while Fig.~\ref{fig:6}(c) depicts the resulting cycle. 

\subsection*{Performance of active engine}
\begin{figure*}
    \centering
    \includegraphics[width=1\linewidth]{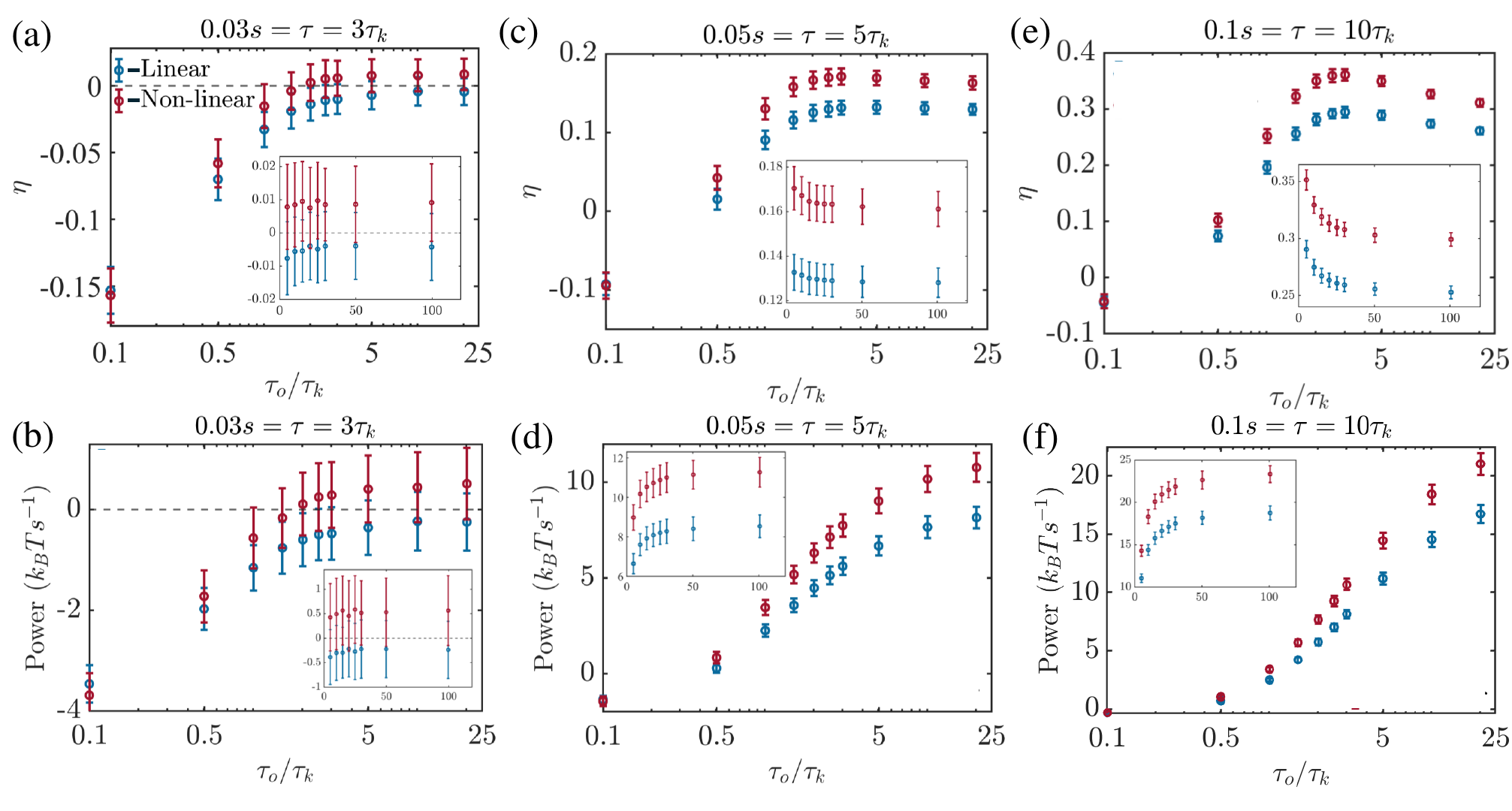}
    \caption{\raggedright Performance plots of the active engine for different cycle times with varying correlation time scales. Plots (a), (c) and (e) show efficiency at cycle times $\tau=0.03s=3 \tau_k$, $\tau=0.05s=5 \tau_k$ and $\tau=0.1s=10 \tau_k$, respectively at short correlation time (insets: efficiency at large correlation time scales). Plots (b), (d) and (f) are the corresponding powers. (insets: power at large correlation time scales)}
    \label{fig:7}
\end{figure*}
Analogous to the passive engine, we evaluate the efficiency and power using trajectories obtained from the numerical integration of Eqs.~\eqref{eq:active_engine} and \eqref{eq:active_drive}, under the prescribed engine protocol. A consistent and substantial enhancement in performance is observed across all combinations of engine cycle times and correlation times for the engine under the nonlinear protocol, as illustrated in the performance plots in Fig.~\ref{fig:7}. Both the cycle times $\tau$ and the correlation times $\tau_o$ are expressed in terms of $\tau_k = \gamma/k_{min}$ for better comparison.

Notably, under certain conditions, our results show a clear transition from negative to positive efficiency (and in power), indicating that the engine, which was previously operating as a heat pump under the standard linear protocol, now functions as a genuine heat engine when driven by the nonlinear protocol (see Fig.~\ref{fig:7}(a), (b), where the cycle time is fixed while the persistence time scale is varied). This is clearly observed when the cycle time is considerably low $\sim 3\tau_k$. Moreover, as further shown in Figs.~\ref{fig:7}(c)-(f), even for higher cycle times $\sim 5\tau_k - 10\tau_k$, over a large range of persistence timescales ($0.1\tau_k \leq \tau_o < 25\tau_k$), the nonlinear protocol consistently delivers better performance as an engine than the linear one.  

\begin{figure*}
    \centering
    \includegraphics[width=1\linewidth]{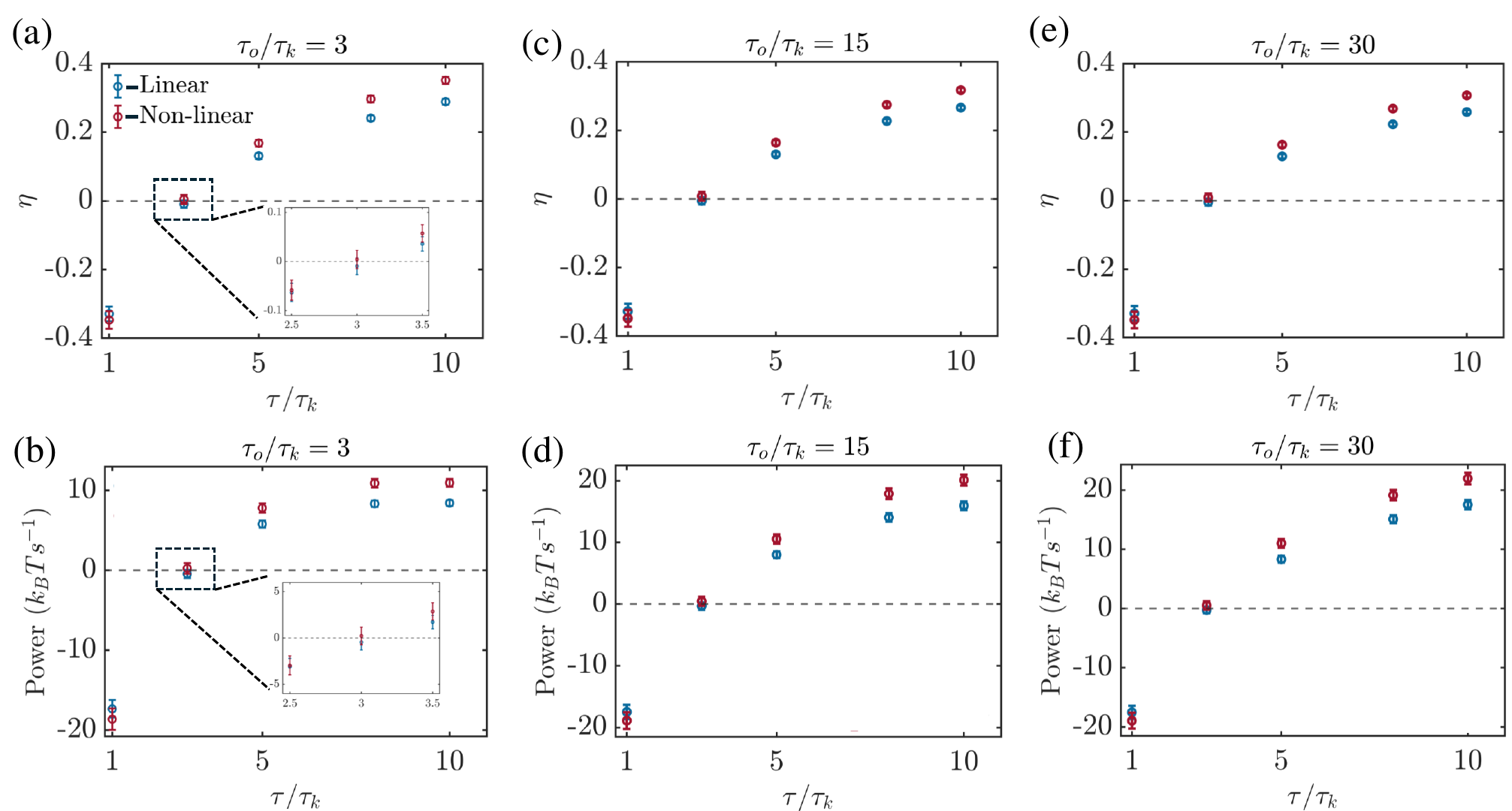}
    \caption{\raggedright Performance plots of the active engine with varying cycle times at different correlation time scales of the active forces. Plots (a), (c) and (e) are efficiency at different correlation times: $\tau_o=3 \tau_k$, $\tau_o=15 \tau_k$ and $\tau_o=30 \tau_k$ respectively, while varying the cycle times and (b), (d) and (f) are the corresponding powers.}
    \label{fig:8}
\end{figure*}

Alternatively, in Figs.~\ref{fig:8}(a)-(f) - where we vary the cycle time from $\tau_k$ to $10\tau_k$, keeping the persistence timescale fixed, we find that the nonlinear protocol enables the system to begin functioning as an engine at a slightly earlier cycle time (as shown in the enlarged part of ~\ref{fig:8}(a) and (b)) than its linear counterpart. Moreover, across the entire range of cycle times considered, the nonlinear protocol consistently delivers higher efficiency and power, regardless of the persistence timescale of the active noise.

\begin{figure}
    \centering
    \includegraphics[width=1\linewidth]{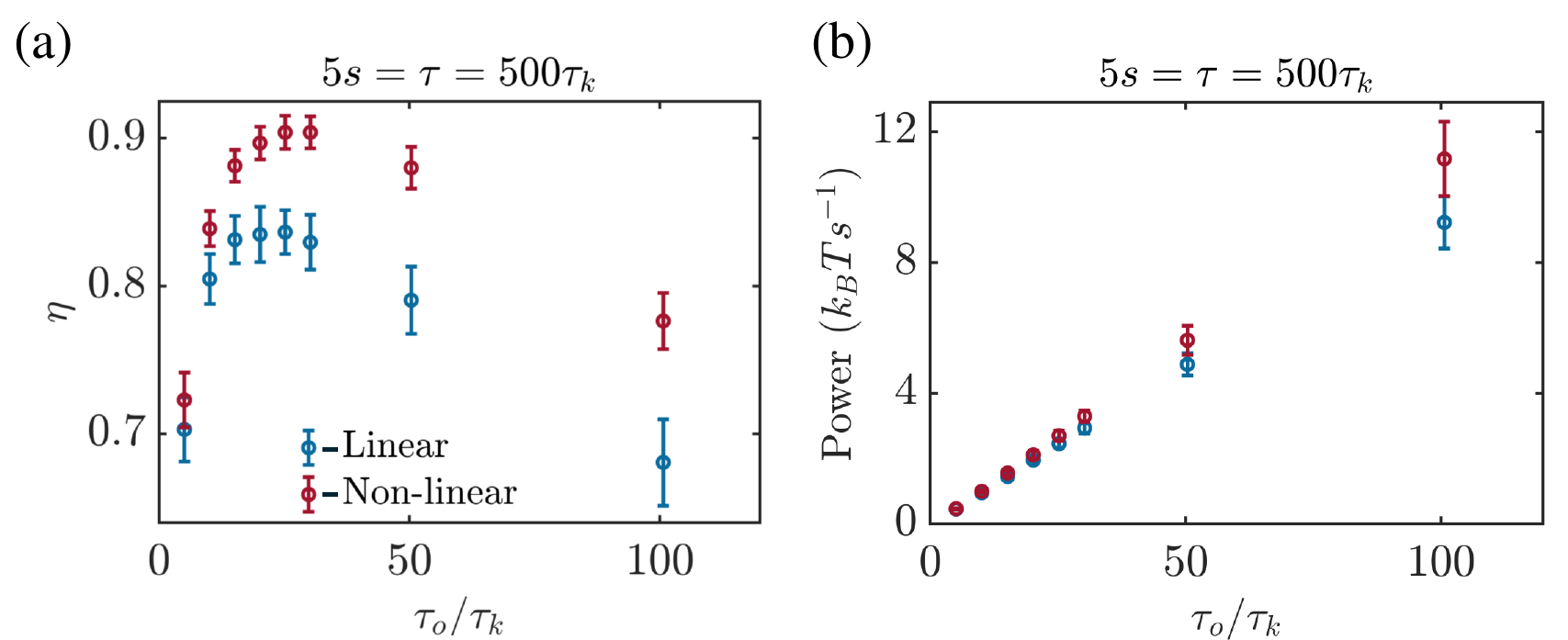}
    \caption{\raggedright Performance plots of the active engine in the quasistatic limit. (a) Efficiency and (b) corresponding power outputs as a function of varying correlation time scales.}
    \label{fig:9}
\end{figure}

An additional remarkable feature is that the active engine with the nonlinear protocol is found to surpass the quasi-static efficiency $\eta_q$ faster for specific combinations of $\tau$ and $\tau_o$. Furthermore, increasing the control parameter $T_A$, which enhances activity and thereby significantly raises the effective temperature of the bath, can lead both the engines to exceed the high-temperature efficiency limit~\cite{saha2019stochastic, albay2023colloidal,roy2021tuning} $\eta_{\infty}=\frac{1}{1+\frac{1}{\ln(\frac{k_{max}}{k_{min}})}}=0.41$ and eventually the Carnot limit, $\eta_c$ (Fig.~\ref{fig:9}) as well. These results clearly demonstrate that the conventional power efficiency trade-off does not necessarily hold for active engines. In contrast to passive engines, where increasing power output typically leads to increased dissipation, and thus enforces a trade-off between power and efficiency, the inherently non-Markovian nature of the active bath provides a constant source of generating energy, thereby suppressing irreversible losses at finite cycle times, which can result in simultaneous realisation of high power and efficiency~\cite{pietzonka2018universal,PhysRevE.102.032116}. 

Finally, in the quasi-static limit (Fig.~\ref{fig:9}), the nonlinear active engine's efficiency approaches the theoretical maximum of unity ($\eta = 1$) in the regime where $\frac{T_c}{T_h} \to 0$, corresponding to a very high effective temperature $T_h$.
This highlights the robustness of the nonlinear protocol in enhancing the performance of active engines when the driving force carries temporal correlations. The improvement persists across a wide range of cycle and correlation times, indicating that it is not restricted to finely tuned conditions. Notably, the nonlinear protocol can transform a heat-pump–like operation into a genuine heat engine and, under strong activity, even surpass both the quasi-static and high-temperature efficiency bounds -- underscoring its effectiveness in realistic non-equilibrium settings (like in the bacterial bath~\cite{majhi2025decoding} or intracellular active media~\cite{eskandari2025active}).

\section{Conclusion}

\begin{figure*}
    \centering
    \includegraphics[width=1\linewidth]{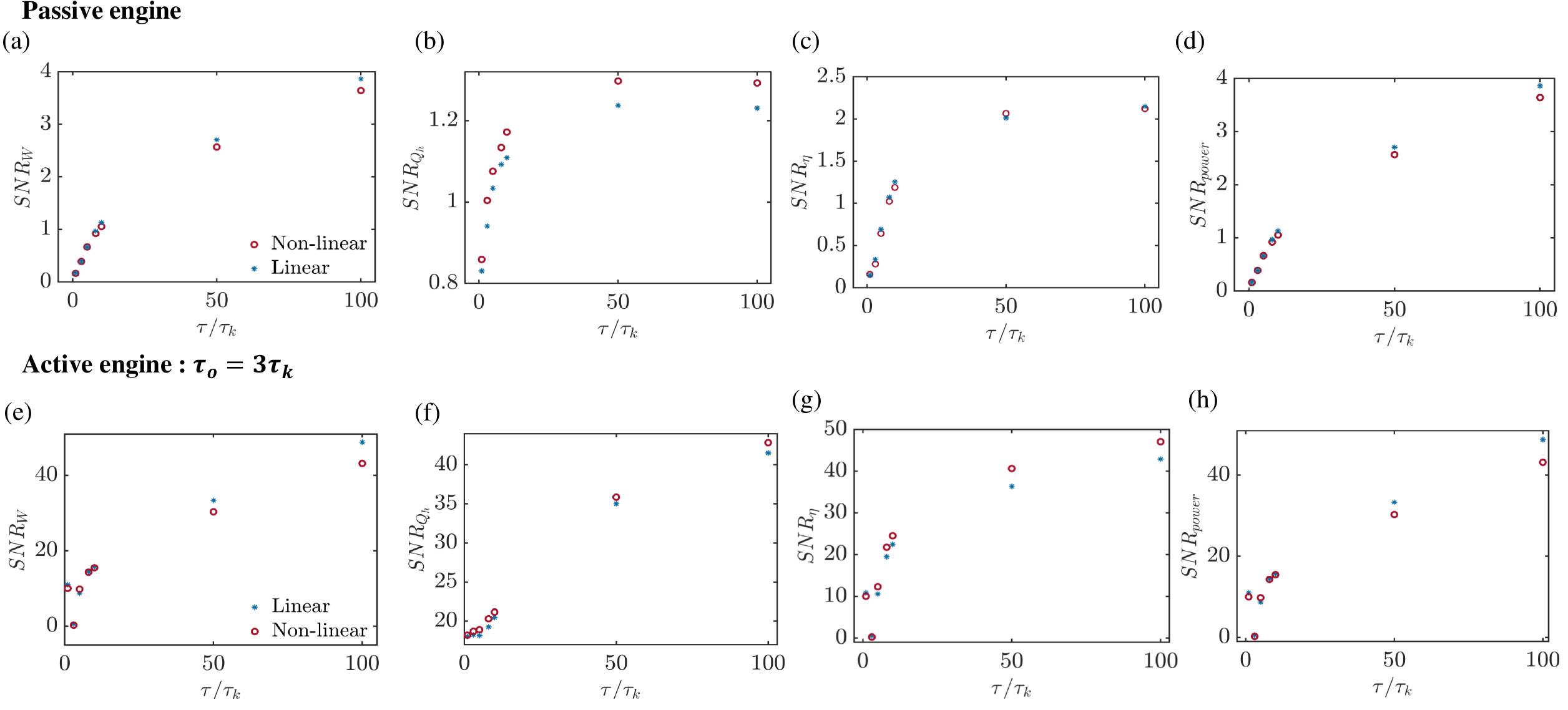}
    \caption{\raggedright Signal to noise ratio for both protocols in the passive engine [panels (a) - (d)] and in the active engine [panels (e) - (h)] at a fixed correlation time of $\tau_o = 3\tau_k$, calculated as the ratio of the square of the mean value to its variance, (e.g., $SNR_{W}=\frac{\langle W \rangle^2}{\mathrm{Var}(W)}$). The subscripts in the y-axis labels indicate the corresponding thermodynamic quantities.}
    \label{fig:10}
\end{figure*}

 We have demonstrated that a single, symmetric cubic stiffness profile provides a powerful, unified driving protocol for colloidal Stirling engines in both passive (thermal) and active (Ornstein–Uhlenbeck) environments.  By closely following quasi-equilibrium paths throughout each cycle, this nonlinear trajectory substantially suppresses irreversible losses and elevates both work output and efficiency under finite-time operation. Notably, it enables the passive engine to exceed the seminal Curzon–Ahlborn bound at maximum power \cite{curzon1975efficiency}, and allows the active engine to transcend its conventional high-temperature limit, while even converting regimes that acted as mere heat pumps under linear driving into genuine engines. To further substantiate this observation, we also present in the Appendix B an alternative family of nonlinear stiffness-modulation protocols that exhibit similar improvements in efficiency and power. Note here, that in the context of mesoscopic thermodynamic processes, nonlinear time-dependent driving protocols commonly emerge when one seeks to optimize or extremize the output of the process, although the specific objective may vary across systems. In some cases, the goal is to maximize average work extraction in cyclic ~\cite{schmiedl2008efficiency} or non-cyclic processes~\cite{schmiedl2007optimal,pal2014extracting}, while in others it is to minimize dissipation. Stochastic thermodynamic “shortcut” protocols, including adiabatic and isothermal shortcuts~\cite{deffner2014classical,PhysRevE.106.054108,zhao2022low,PhysRevE.96.012144,martinez2016engineered,guery2019shortcuts,guery2023driving}, fall into this class, where dissipation is explicitly optimized or controlled.

However, importantly, a fully optimal protocol for a finite-time cyclic engine needs to be constrained by the coexistence of multiple competing objectives -- maximizing work or power, minimizing dissipation, and maintaining stability against fluctuations -- while simultaneously driving both trap stiffness and temperature ~\cite{schmiedl2007optimal,schmiedl2007efficiency,chatterjee2025optimal,PhysRevE.98.022133,ye2022optimal}. In active environments, these challenges are further compounded by additional non-equilibrium timescales and effective temperatures, which are not known a priori and may vary during the cycle. As a result, analytically tractable optimal protocols are difficult to construct and may lack robustness against parameter uncertainty. Moreover, optimal control solutions often rely on exact knowledge of system parameters and may require sharply varying or discontinuous control fields, which are difficult to implement experimentally in colloidal systems. Nevertheless, our results provide intuitive guidance toward optimal driving: it appears that nonlinear protocols shape the evolution in a way that suppresses irreversible contributions throughout the cycle leading to a consistent improvement in finite-time engine performance, thus indicating a practical pathway toward near-optimal operation. In particular, the nonlinear protocols appear to suppress the finite-time relaxation lag of the system from reaching an instantaneous equilibrium distribution, thereby improving engine performance at short cycle times. This suggests that near-optimal protocols should allocate more time to regions of strong sensitivity to stiffness or temperature changes while accelerating the dynamics where relaxation is faster, thereby ensuring that the engine maintains a somewhat quasistatic characteristic.
Crucially, our protocol requires no exotic components or real-time feedback~\cite{albay2023colloidal}, but only time-programmed modulation of standard trap stiffness and noise amplitude, thereby making it fully compatible with existing optical tweezers platforms.  This should bridge the longstanding gap between theoretical cycle optimisation and practical implementation, pointing the way toward fast, high-power, and high-efficiency microscale heat engines for lab-on-a-chip energy harvesting~\cite{wu2022perspective}, nanoscale sensing and beyond~\cite{liu2013small,yue2012nanoscale,ebbens2010pursuit,patel2006nanorobot,douglas2012logic}. Looking ahead, experimental validation on optical-tweezers setups, exploration of multi-step or feedback-enhanced cycles, and extension to other mesoscopic engines will further solidify the role of nonlinear driving in next-generation stochastic thermodynamic devices.

\section*{DATA AVAILABILITY}
The data that support the findings of this study are available upon reasonable request from the authors.

\begin{acknowledgments}
The work is supported by IISER Kolkata and the Science and Engineering Research Board (SERB), Department of Science and Technology, Government of India, through the research grant CRG/2022/002417. A.S. acknowledges Department of Science and Technology, Govt. of India for the ARG-ANRF grant (file number: ANRF/ARG/2025/009343/PS). A.S. also acknowledges the funding ({\it{Investissements d’Avenir}}, ANR-16-IDEX-0008) by CY Initiative of Excellence. This work was partially developed during his stay at the CY Advanced Studies, whose support is acknowledged. BD is thankful to the Ministry Of Education of Government of India for financial support through the Prime Minister’s Research Fellowship (PMRF) grant.
\end{acknowledgments}

\appendix
\section{\label{A} Signal to Noise plots for both passive and active engines}
\begin{figure}
    \centering
    \includegraphics[width=1\linewidth]{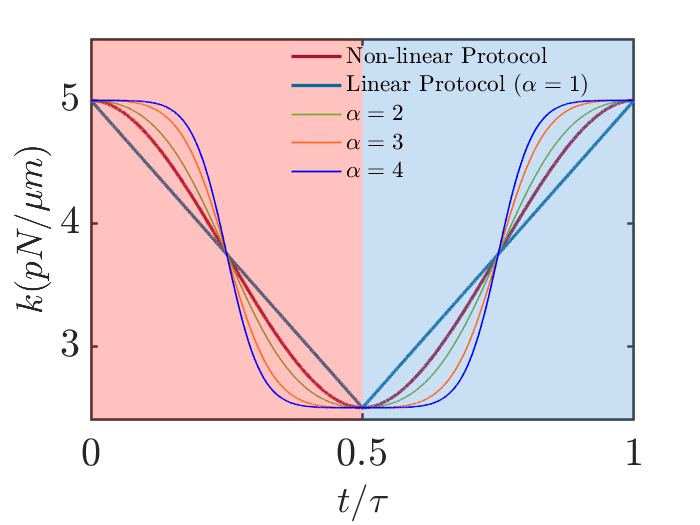}
    \caption{\raggedright Alternative stiffness modulation protocol with tunable nonlinearity parameter $\alpha$, where larger $\alpha$ corresponds to stronger non-linearity. }
    \label{fig:11}
\end{figure}
\begin{figure}
    \centering
    \includegraphics[width=1\linewidth]{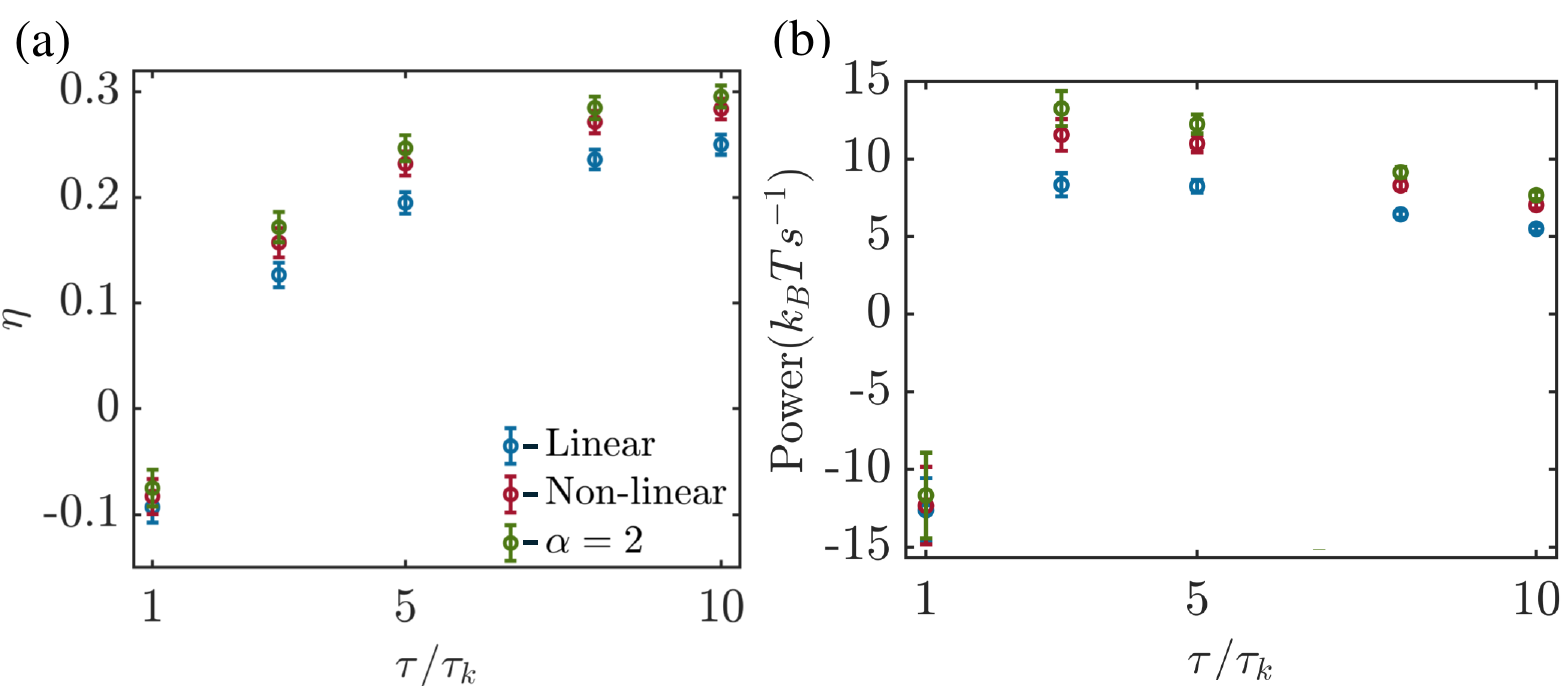}
    \caption{\raggedright Passive engine performance at varying cycle times, comparing the linear protocol ($\alpha=1$)(blue) and non-linear protocol(red) from the main text with an alternative stiffness modulation protocol with $\alpha=2$(green).}
    \label{fig:12}
\end{figure}
In assessing performance-related fluctuations, we observed that the non-linear driving does not inherently lead to an improved signal-to-noise ratio which was calculated as the ratio of the square of the mean value of the specific thermodynamic quantity to its variance as can be seen from Fig.~\ref{fig:10}, nor does enhanced performance require increased fluctuations. In our simulations, switching from linear to nonlinear protocols does not produce a noticeable change in the fluctuations of thermodynamic observables. Instead, the performance improvement primarily stems from a reduction in the mean irreversible work, while the variances remain largely unchanged. Thus, the enhanced efficiency and power arise from lowering the dissipation parameter $\Sigma$, rather than from exploiting larger fluctuations.
\

\section{\label{B} Alternative non-linear protocol}
We tested another family of non-linear curves (Fig.~\ref{fig:11}), defined by the following equation as driving protocols for both passive and active engines.
\begin{equation}
    k(t)=\begin{cases}
        k_{max}-\Delta k \frac{(s)^\alpha}{(s)^\alpha + (1-s)^\alpha},&  0\leq s\leq 1\\
        k_{max}-\Delta k \frac{((2-s))^\alpha}{((2-s))^\alpha + (1-(2-s))^\alpha},&  1< s\leq 2
    \end{cases}
\end{equation}
where $s=\frac{2t}{\tau}$ and $\alpha$ is a control parameter determining the degree of nonlinearity in the stiffness variation.\\
For the passive engine, we have plotted the performance corresponding to $\alpha =2$ curve, since only for this we could maintain the effective temperature at 1000 K by varying $\theta$ within the range 0.45 to 2.0. The case $\alpha=1$ corresponds to the linear protocol discussed in the main text.

\begin{figure}
    \centering
    \includegraphics[width=1\linewidth]{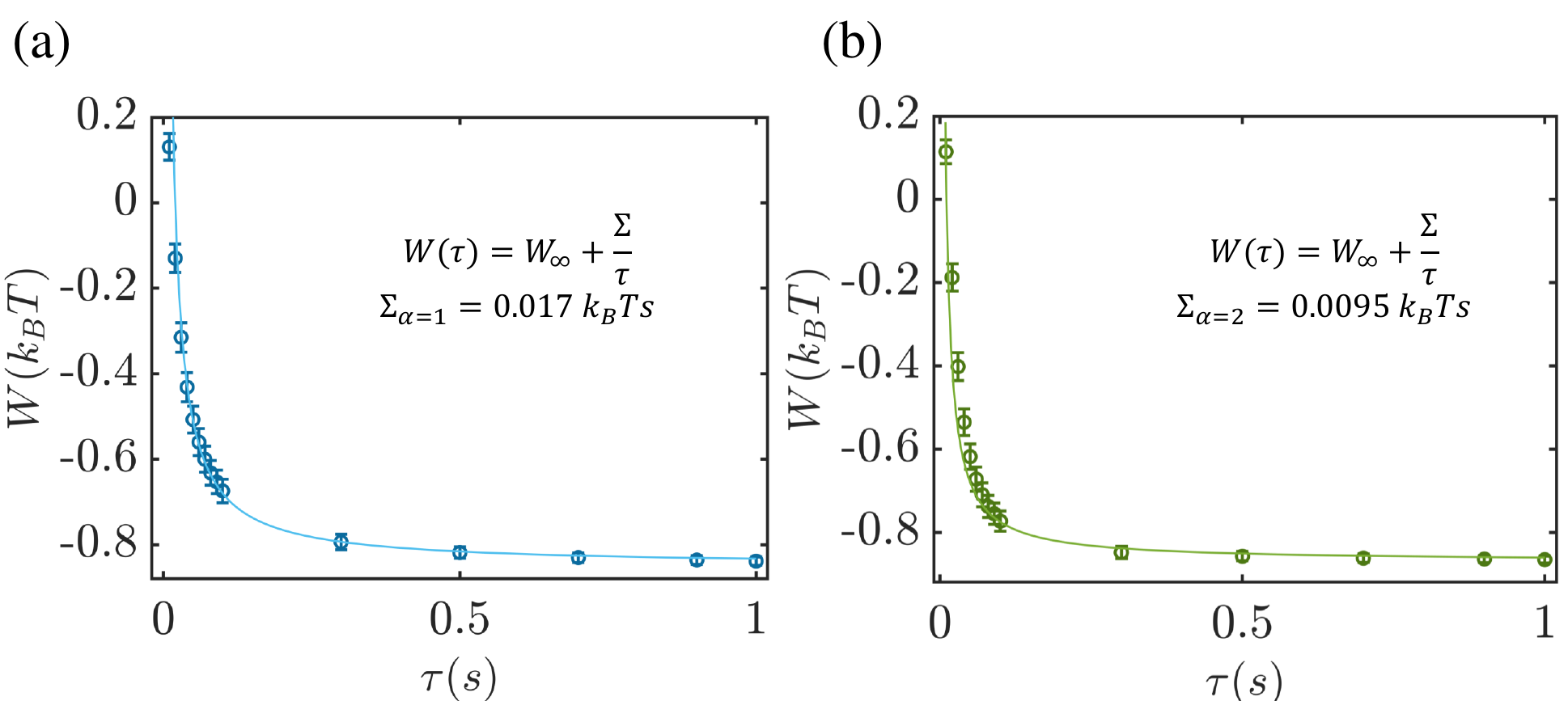}
    \caption{\raggedright Mean work done in the passive engine under both (a) $\alpha=1$(linear) and (b) $\alpha=2$. The analytical expression shows good agreement with simulation data. The fitted irreversibility parameters are $0.017 k_B T s$ for the $\alpha=1$ protocol matching the value obtained for the linear protocol in the main text and $0.0095 k_B T s$ for the $\alpha=2$ protocol.}
    \label{fig:13}
\end{figure}

The corresponding performance plots for $\alpha = 2$ are shown in (Fig.~\ref{fig:12}). We observe a similar enhancement in performance compared to the linear protocol, consistent with the trends obtained for the previously discussed non-linear protocol in the main text.

we analyzed the mean work done for $\alpha=1$ (which is the same as the linear protocol) and $\alpha=2$ by fitting the data to $W(\tau)=W_{\infty}+W_{diss} \equiv W_{\infty}+\frac{\Sigma}{\tau}$. For $\alpha=1$ the irreversibility parameter is identical to that obtained for the linear protocol. In contrast, for $\alpha=2$ the irreversibility parameter is significantly reduced, with $\Sigma_{\alpha=2}=0.0095 k_B Ts$ (Fig.~\ref{fig:13}) indicating lower dissipation.

\begin{figure}
    \centering
    \includegraphics[width=1\linewidth]{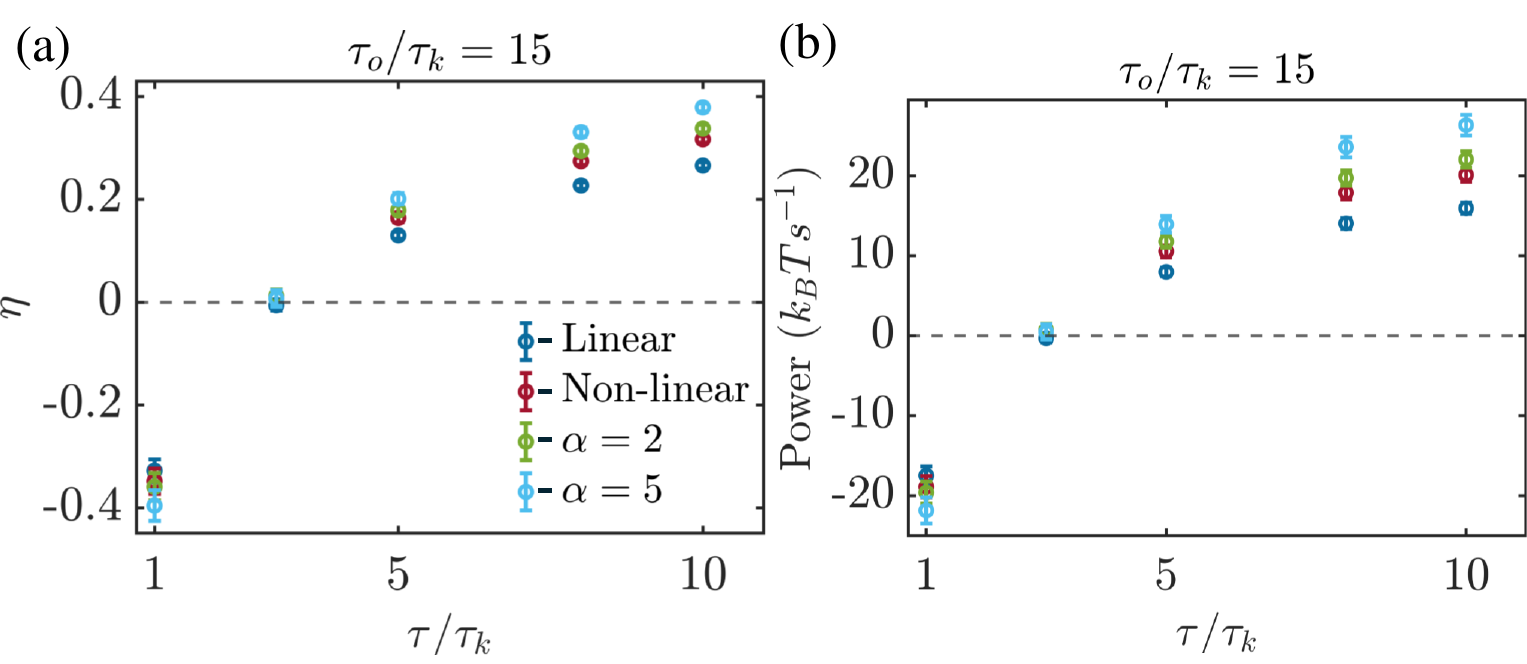}
    \caption{\raggedright Performance plots for active engine comparing the linear protocol ($\alpha=1$) (blue), non-linear (red), both as discussed in the main text and the alternative stiffness modulation protocols for $\alpha=2$ (green) and $\alpha=5$ (cyan).}
    \label{fig:14}
\end{figure}
We further applied this new family of driving protocols to the active engine.  The performance is shown for one specific case in which the cycle time is varied at a fixed correlation time of $\tau_o=15 \tau_k$ for different values of the non-linearity parameter $\alpha$ (Fig.~\ref{fig:14}). We observe an clear increase in performance with increasing degree of non-linearity. From our observations, we can argue that non-linear driving protocols which reduce the dissipated work quantified by a lower irreversibility parameter $\Sigma$ are beneficial for designing a more efficient microscopic heat engine.

\section{\label{C} Analytical evaluation of lag and irreversible work}
The improved performance of the non-linear protocol can be understood by analyzing the lag between the actual state of the trapped particle and the instantaneous equilibrium distribution associated with the time-dependent trap stiffness. Since the irreversible work (dissipation) generated during finite-time driving originates from this lag, directly comparing this lag across protocols gives insight into the observed reduction in dissipation for the non-linear engine.\\
The dynamics of an overdamped Brownian particle trapped in a harmonic trap with time-dependent stiffness $k(t)$ can be written in the following form as -

\begin{equation}
\label{eq_apC_langevin}
    \gamma \dot x(t) = -k(t) x(t) + \xi(t)
\end{equation}

where the noise satisfies: $\langle \xi(t)\xi(t') \rangle = 2\gamma k_B T_{eff}\delta(t-t')$ with $\langle \xi(t) \rangle =0$. Note that $T_{eff}$ can be controlled through the application of external noise to the trap center as mentioned in the main text.  The instantaneous variance is defined as $\sigma (t)=\langle x^2(t)\rangle$, assuming $\langle x(t)\rangle =0$. Using Ito calculus, the temporal dynamics of $x^2$ can be written as:
\begin{equation}
    \frac{d(x^2)}{dt}=
    2x\frac{dx}{dt} +
    \underbrace{\frac{2k_BT_{eff}}{\gamma}}_{\text{Ito correction}}
\end{equation}
Then substituting the Langevin equation (Eq.\eqref{eq_apC_langevin}):

\begin{equation}
   \frac{d(x^2)}{dt}=2x\Bigl[-\frac{k(t)}{\gamma}x(t)+\frac{\xi(t)}{\gamma}\Bigr]+\frac{2k_BT_{eff}}{\gamma}
\end{equation}
Taking the ensemble average of the above leads to:
\begin{equation}
    \frac{d\langle x^2 \rangle}{dt}=-\frac{2k(t)}{\gamma}\langle x^2 \rangle+\frac{2k_BT_{eff}}{\gamma}
\end{equation}
Therefore, the temporal dynamics of the instantaneous  variance can be written as, 
\begin{equation}
\label{eq_apC_variance}
    \frac{d\sigma (t)}{dt}=-\frac{2k(t)}{\gamma}\sigma (t)+\frac{2k_BT_{eff}}{\gamma}
\end{equation}

\begin{figure}
    \centering
    \includegraphics[width=1\linewidth]{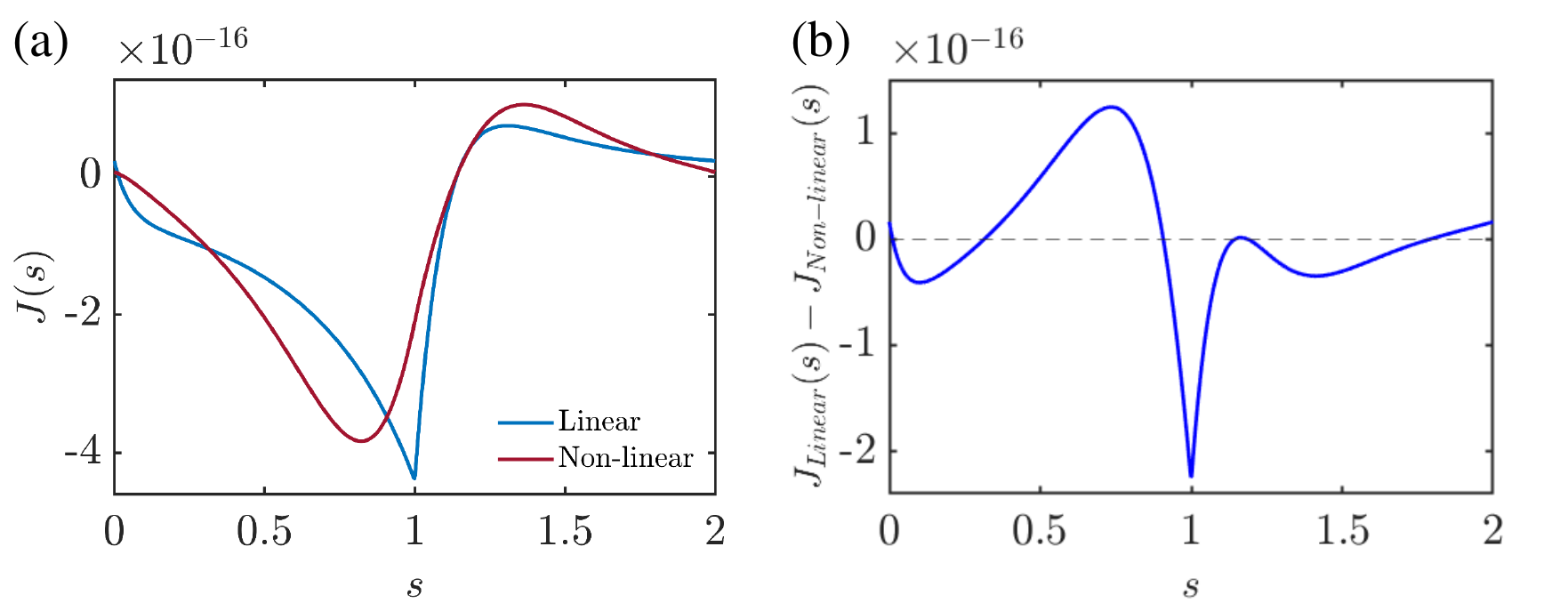}
    \caption{\raggedright Analytical lag function. (a) Periodic steady-state lag $J(s)$ for linear and non-linear protocol. (b) Difference in lag, $J_{linear}(s)-J_{non-linear}(s)$, for one period.}
    \label{fig:15}
\end{figure}

Note that for an instantaneous value of the trap stiffness ($k(t)$),  the corresponding equilibrium variance is: $\sigma_{eq}(t)=\frac{k_BT_{eff}}{k(t)}$. In the limit of infinitely slow driving, the system remains in quasiequilibrium throughout the process, with $\sigma(t)=\sigma_{eq}(t)$. In contrast, finite driving causes a delayed response to the instantaneous equilibrium state of the system. 

To quantify the effect of the delayed response on the engine performance, we characterize the system dynamics over a complete cycle by introducing the lag variable $J(t)=\sigma(t)-\sigma_{eq}(t)$, which measures the deviation of the instantaneous variance from its equilibrium value.

Differentiating this quantity
\begin{figure}
    \centering
    \includegraphics[width=1\linewidth]{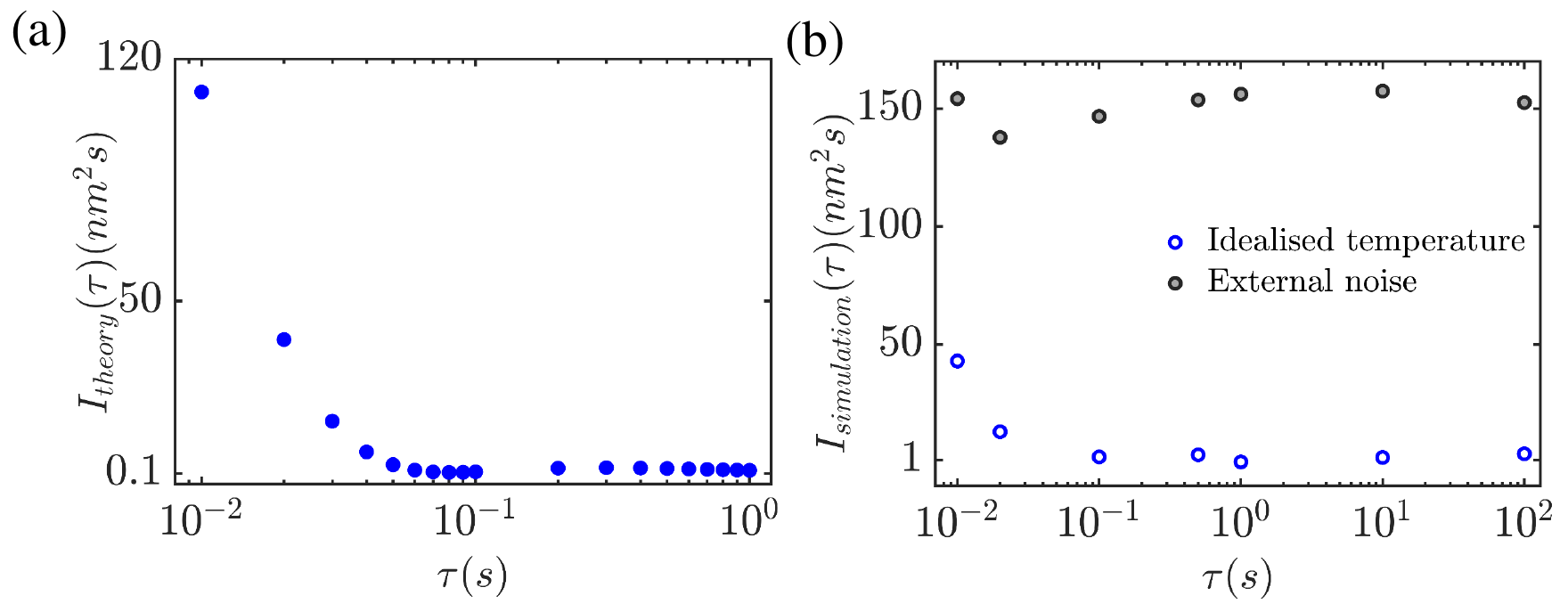}
    \caption{\raggedright Integrated lag $I(\tau)$ as a function of cycle time $\tau$ for (a) analytical estimation and (b) numerical simulations of the engine. In (b), results are shown for an idealised isothermal temperature profile (blue hollow circles) and for the externally driven system with noise (black filled circles). In all cases, ($I(\tau) >0$) for all cycle times, indicating that the non-linear protocol exhibits a smaller lag and remaining closer to equilibrium compared to the linear protocol.}
    \label{fig:16}
\end{figure}
\begin{equation}
    \frac{dJ}{dt}=\frac{d\sigma}{dt}-\frac{d\sigma_{eq}}{dt}
\end{equation}
Substituting the dynamical equation of variance (Eq.\eqref{eq_apC_variance}) into this gives:
\begin{equation}
    \frac{dJ}{dt}=-\frac{2k(t)}{\gamma}\sigma+\frac{2k_BT_{eff}}{\gamma}-\frac{d\sigma_{eq}}{dt}
\end{equation}
Since $\sigma_{eq}=\frac{k_BT_{eff}}{k(t)}$, then $\frac{d\sigma_{eq}}{dt}=-\frac{k_BT_{eff}}{k^2}\frac{dk}{dt}$. Substituting $\sigma=J(t)+\sigma_{eq}$:
\begin{equation}
    \frac{dJ}{dt}=-\frac{2k}{\gamma}(J(t)+\sigma_{eq})+\frac{2k_BT_{eff}}{\gamma}+\frac{k_BT_{eff}}{k^2}\frac{dk}{dt}
\end{equation}
Since $-\frac{2k}{\gamma}\sigma_{eq}+\frac{2k_BT_{eff}}{\gamma}=0$, we obtain the exact lag equation for any arbitrary $k(t)$:
\begin{equation}
\label{eq_apC_lagdynamics}
    \frac{dJ}{dt}=-\frac{2k}{\gamma}J+\frac{k_BT_{eff}}{k^2}\frac{dk}{dt}
\end{equation}
Introducing dimensionless time $s=\frac{2t}{\tau}$, where $\tau$ is the engine cycle time, Eq.~\eqref{eq_apC_lagdynamics} becomes:
\begin{equation}
    \frac{dJ (s)}{ds}+\frac{\tau k(s)}{\gamma}J(s)=\frac{k_BT_{eff}}{k(s)^2}\frac{dk(s)}{ds}
\end{equation}
Note that $J(s)$ obeys the first-order linear ODE and for the periodic Stirling cycle, $J(s+2)=J(s)$, which uniquely determines the periodic steady-state solution. Using the integrating factor,
\begin{equation}
    \mu(s)=\exp \Bigl[\frac{\tau}{\gamma}\int_{0}^{s}k(s')ds' \Bigr]
\end{equation}
The solution of the ODE is:
\begin{equation}
    J(s)=\mu(s)^{-1} \Bigl[C+\int_{0}^{s}\mu(s')\frac{k_BT_{eff}}{k(s')^2}\frac{dk(s')}{ds'}ds' \Bigr]
\end{equation}
where the constant $C$ is determined as $C=\frac{\int_{0}^{P}\mu^{-1}\frac{k_BT_{eff}}{k(s')^2}\frac{dk(s')}{ds'}ds'}{\mu(P)-1}$, using the periodicity condition $J(0)=J(2)$, (period, $P=2$). \\

We evaluated this analytical solution for engine with linear and nonlinear variations in stiffness (as given in Eq.\eqref{eq_linear_kt} and Eq.~\eqref{eq_nonlinear_kt} in the main text, respectively), in each isotherms using idealised temperature profiles with a hot isotherm at $1000 K$ and a cold isotherm at $295 K$. The resulting steady state lag $J(s)$ over one engine cycle is shown in Fig.~\ref{fig:15}(a). The difference in lag between the protocols was quantified by computing $J_{diff}=J_{Linear}(s)-J_{Non-linear}(s)$ and its integrated value over a cycle $I(\tau)=\int_{0}^{P}J_{diff}(s)ds$. The corresponding difference $J_{diff}(s)$ is plotted in Fig.~\ref{fig:15}(b). 
\begin{figure}
    \centering
    \includegraphics[width=1\linewidth]{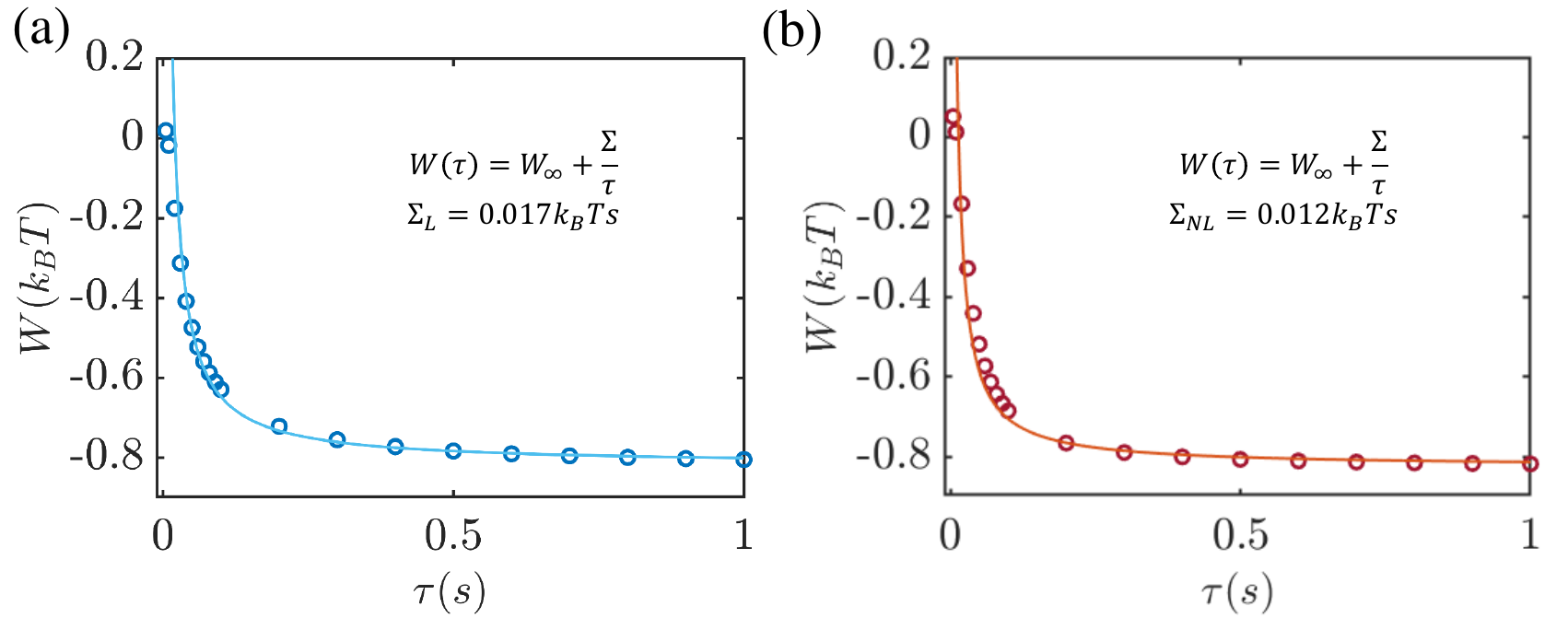}
    \caption{\raggedright Average work (symbols) calculated from analytical variance as a function of cycle duration for (a) linear and (b) non-linear protocol. Solid lines are the fits to the equation $W(\tau)=W_{\infty}+\frac{\Sigma}{\tau}$.}
    \label{fig:17}
\end{figure}
For all cycle durations investigated, $I(\tau)>0$ as observed in Fig.~\ref{fig:16}. 
Panel (a) of Fig.~\ref{fig:16} shows the analytical calculation for the engine with an idealised isothermal
temperature profile fixed a priori, whereas panel (b) corresponds to the numerical 
simulation of the same engine protocol. The results demonstrate that the non-linear protocol remains systematically closer to equilibrium compared to the linear protocol as $I(\tau) >0$.%

The reduction in lag should correspond to a reduction in irreversible work. To verify 
this analytically, we calculate the average work per cycle from the periodic 
steady-state variance as
\begin{equation}
    W=\frac{1}{2}\oint \sigma(s)\frac{dk}{ds}ds .
\end{equation}
The analytical estimates are shown in Fig.~\ref{fig:17} along with the Sekimoto-Sasa fit. The fitted irreversibility parameters are $\Sigma_L=0.017k_BT$ for the linear 
protocol and $\Sigma_{NL}=0.012k_BT$ for the non-linear protocol, confirming the 
reduction in dissipation. These values are consistent with those obtained from 
stochastic simulations shown in Fig.~\ref{fig:4} of the main text.


\bibliography{apssamp}

\end{document}